\documentclass[reqno]{amsart}
\usepackage{psfrag}
\usepackage{color}
\usepackage{amsmath}
\usepackage{amssymb}
\usepackage{graphicx}
\usepackage{natbib}

\begin{document}
\title[Dissolution from forced convection]{Impact of background flow on
dissolution trapping of carbon dioxide injected into saline aquifers}
\author{Saikiran Rapaka}
\address{Computational Earth Sciences Group, Los Alamos National Laboratory,
Los Alamos NM 87544}
\email{sai@lanl.gov}

\author{Rajesh J. Pawar}
\address{Computational Earth Sciences Group, Los Alamos National Laboratory, 
Los Alamos NM 87544}
\email{rajesh@lanl.gov}

%\author{Rajesh J. Pawar}
%\affiliation{EES-16, Los Alamos National Laboratory, Los Alamos NM 87544}
\date{}

\begin{abstract}
While there has been a large interest in studying the role of dissolution-driven
free convection in the context of geological sequestration, the contribution of
forced convection has been largely ignored. This manuscript considers CO$_2$
sequestration in saline aquifers with natural background flow and uses
theoretical arguments to compute the critical background velocity needed to
establish the forced convective regime. The theoretical arguments are supported
by two dimensional high-resolution numerical simulations which demonstrate the
importance of forced convection in enhancing dissolution in aquifers
characterised by low Rayleigh numbers.
\end{abstract}

%\begin{document}
\maketitle
\section{Introduction}
Geological storage of carbon dioxide in deep saline aquifers is considered to be
one of the key technological solutions to mitigating carbon emissions in the
near future (\cite{IPCCreport}). Under the thermodynamic conditions at which
CO$_2$ is injected into subsurface formations, CO$_2$ exists in a supercritical state with a density that is lower
than that of brine in the medium. Due to this difference in density, CO$_2$
rises to the top of the domain and containment is achieved by the presence of
an overlying low-permeability caprock. 

Due to the tremendous importance of understanding the processes leading to
storage security of injected carbon dioxide, there has been a large body of
published literature exploring the different trapping mechanisms. The physical
processes that play a role at increasing timescales are $(i)$ physical trapping due to the caprock, $(ii)$
residual trapping of bubbles of CO$_2$ in the porous rock, $(iii)$ dissolution
trapping due to the slow dissolution of CO$_2$ into the brine and $(iv)$ mineral
trapping due to chemical reactions between dissoved carbon dioxide and host
rocks. The work presented in this manuscript is concerned with dissolution
trapping mechanism under different aquifer flow conditions. Experimental data
has shown that the dissolution of CO$_2$ into brine produces a weak density
contrast, with CO$_2$ saturated brine being slightly denser than unsaturated brine. Most
recent work in this area has focused on the free convective mixing that can
arise due to this difference in density. 

One of the effects that has so far been neglected is the contribution of
background flow to the rate of dissolution of CO$_2$ into the brine in the
aquifer. The only exception we are aware of is the work by
\cite{HassanBackground}, where the authors have analysed the role of additional
dispersion due to the background flow. For a finite sized CO$_2$ plume in the
supercritical state, the background flow leads to the development of a boundary
layer in which horizontal advection balances the diffusion of dissolved carbon
dioxide. We will henceforth refer to the dissolution occuring due to the
background flow as forced convection. A compilation of reservoir characteristics
for potential sites in western Canada by \cite{BachuData} suggests that the
Rayleigh numbers are going to be fairly low for a large number of storage sites.
Recent computations (\cite{Rapaka}) as well as experiments (\cite{Backhaus}) in
analogous systems have shown that the free convective mechanism (fingering) can
take very long to establish under such conditions. It is possible that forced
convection can play the dominant role in enhancing storage security in this
scanario. However, in the context of geological storage of carbon dioxide, the
conditions under which forced convection is the preferred state have not yet
been explored. In this manuscript, we will provide simple theoretical arguments
for the background flow conditions under which forced convection establishes
itself and support these predictions with high resolution numerical simulations.

\section{Governing Equations and Solution Procedure}
We consider the dynamics of CO$_2$ dissolution and transport in a aquifer of
depth $H^{*}$ with an overlying source of CO$_2$ of length $L^{*}$. We assume
the presence of a horizontal background aquifer flow $\mathbf{u}^{*}_0 = (u^{*}_0, 0)$, the
magnitude of which determines whether the system is in a free or forced convective regime. 

The equations needed to describe the system are those describing conservation of
mass, momentum (Darcy's law) and an advection-diffusion equation for the
concentration of dissolved CO$_2$. Experimental data has shown that the density
of brine increases linearly with the concentration of dissolved CO$_2$, with a
maximum increase (under saturated conditions) of the order of 1\%
(\cite{PruessSpycher}). Hence, we utilize the Boussinesq model, where the
density of brine is taken to be a constant everywhere except in the buoyancy
term. The governing equations are:
\begin{align}
\frac{\mu^{*}}{k^{*}} \mathbf{u}^{*} &= -\nabla^{*} P^{*} + \rho_{f}^{*}g
\mathbf{e}_y,\\ 
\nabla^{*}\cdot\mathbf{u}^{*} &= 0,\\
\frac{\partial (\phi C^{*})}{\partial t^{*}} +
\nabla^{*}\cdot\left(\mathbf{u}^{*}C^{*}\right) &= \nabla^{*}\cdot\phi D^{*}
\nabla^{*} C^{*},\\ 
\rho_f^{*} &= \rho_0^{*} + \Delta\rho^{*}\left(\frac{C^{*}}{C_0^{*}}\right),
\end{align}
where, $\mathbf{u}^{*} = \left(u^{*}, v^{*}\right)$ is the velocity of the
fluid, $C^{*}$ is the concentration of dissolved carbon dioxide, $P^{*}$ is the
fluid pressure, $g$ is the acceleration due to gravity, $k^{*}$ is the
permeability of the rock, $\rho_0^*$ is the density of unsaturated brine,
$\Delta\rho^*$ is the density difference between brine saturated with CO$_2$ and
pure brine, $\mu^*$ is the viscosity of brine, $\phi$ is the porosity of the
medium and $D^*$ is the diffusivity of CO$_2$ in brine.

\begin{figure}
\begin{center}
\includegraphics[width = 3.3in]{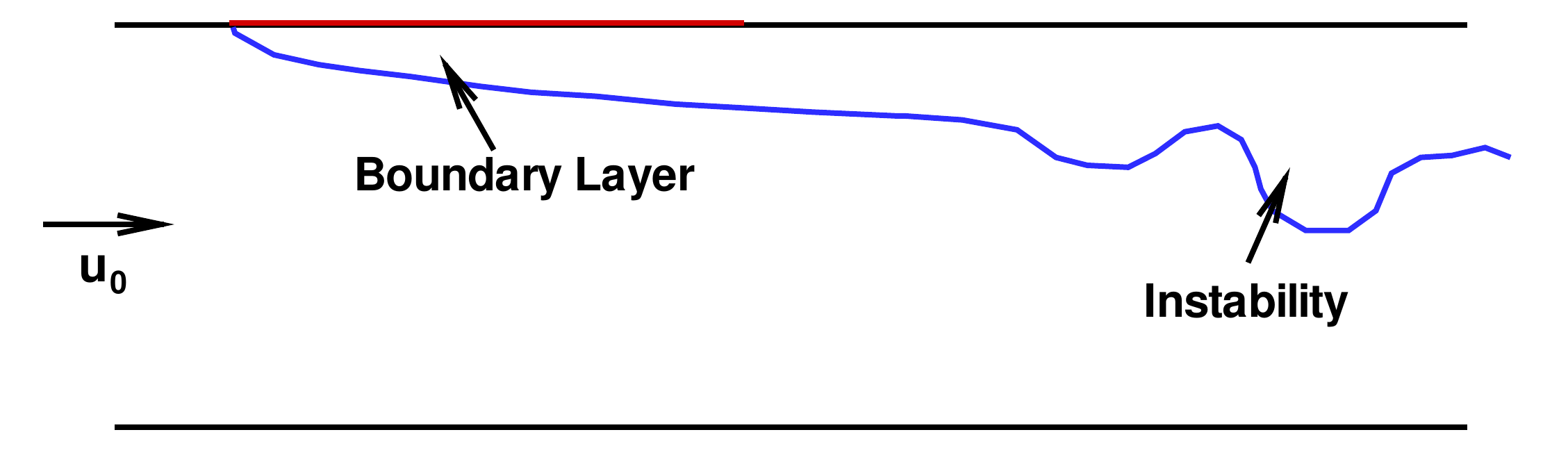}
\vspace{-0.5cm}
\end{center}
\caption{Schematic illustration of the problem geometry considered in this
manuscript. The red part of the top boundary denotes the footprint of CO$_2$
plume in supercritical state. A mean horizontal flow of magnitude $u_0$ is
taken to be flowing through the system.}
 \label{fig:schematic}
\end{figure}

We now non-dimensionalize the equations choosing a convective velocity scale
$U_{r} = k^*\Delta\rho^* g/\mu^*$, the system depth as a length scale $L_{r} =
H^{*}$, and a convective time scale $T_r = L_r/U_r$. Concentrations are
non-dimensionalized using $C_0^*$, the maximum solubility of CO$_2$ in brine at
the conditions in the aquifer. The non-dimensional equations obtained are:
\begin{align}
\mathbf{u} &= -\nabla P + C \mathbf{e}_y \label{eqn:Darcy}\\
\nabla\cdot\mathbf{u} &= 0 \label{eqn:mass}\\
\frac{\partial C}{\partial t} + \nabla\cdot\left(\mathbf{u}C\right) &=
\frac{1}{Ra}\nabla^{2}C \label{eqn:ade}
\end{align}
The non-dimensional paramerer $Ra$ is the ratio of the convective flux to the
diffusive one:
\begin{align}
Ra &= \frac{k^*\Delta\rho^* g H^*}{\mu^*\phi D^*}
\end{align}
The initial conditions are given by $C = 0, u = u_0$ and $v = 0$ everywhere. The
CO$_2$ concentration is fixed on the top boundary to be $C = 1$ for a source
region of length $L$, and a no-flux boundary condition is applied
for the remainder of the top boundary and the entire bottom boudary. On the left boundary, we have
$C = 0, u = u_0$ and $v = 0$ and an outflow boundary condition is applied on the
right boundary. A schematic of the geometry considered is shown in Fig.
(\ref{fig:schematic}). 
% \begin{align}
% C(\mathbf{x}, t=0) = 0, \qquad & \mathbf{u}(\mathbf{x}, t=0) = \left(u_0,
% 0\right),\\
% C(a\leq x\leq a+L, y=0, t) = 1, \qquad & \frac{\partial C}{\partial
% y}(x<a, y, t) = \frac{\partial C}{\partial y}(x>a+L, y, t) = 0\\
% \frac{\partial C}{\partial y}(x, y=1, t) = 0, \qquad & C(x=0, y, t) = 0,\\
% \frac{\partial C}{\partial x}(x = A, y, t) = 0
% \end{align}

Depending on the magnitude of the background flow $u_0$, CO$_2$ dissolution can
be enhanced by free convection (low $u_0$, sufficiently high $Ra$), forced
convection (high $u_0$) or be a purely diffusion-driven process (low $u_0$, low
$Ra$). The case of free convection (especially in the absence of any background
flow) has received a great amount of attention in the recent past (see for
instance \cite{EnnisKing}, \cite{Xu}, \cite{Riaz}, \cite{HassanScaling},
\cite{Rapaka} and \cite{SlimRamakrishnan}). There is an extensive body of
published research on forced convection in porous media over the last four
decades, and it is not possible to present present an overview of even the most
important contributions within the scope of this paper. The reader is referred
to the book by \cite{NieldBejan} where a comprehensive account of the existing
body of work is presented. However, in the context of geological storage of
carbon dioxide, we are not aware of any manuscript that points out the
conditions under which forced convection plays a dominant role in enhancing the
dissolution of injected CO$_2$.
% The streamfunction $\psi$ is defined using $\mathbf{u} =
% \nabla\times\psi\mathbf{e_{z}}$. Taking a curl of Eqn.(\ref{eqn:Darcy}), we
% obtain the following Poisson's equation for the streamfunction: \begin{align}
% \nabla^{2}\psi &= -\frac{\partial C}{\partial x} \end{align}
% \subsection{Forced convection} \cite{NieldBejan}
For sufficiently large background flow rates, we expect to see a forced
convective regime achieving a steady-state dissolution rate. Using the commonly
made assumption of $\frac{\partial^2 C}{\partial x^2} << \frac{\partial^2
C}{\partial y^2}$ in the boundary layer and assuming that the basic velocity
field is of the form $u=u_0, v=0$, the steady state advection diffusion equation
can be reduced to:
\begin{align}
u_{0}\frac{\partial C}{\partial x} &= \frac{1}{Ra}\frac{\partial^2 C}{\partial
y^2}
\end{align}
Looking for similarity solutions of the form $1-C(x,y) = f(\xi)$ in terms of the
similarity variable $\xi = y/\sqrt{x}$, it is easy to show that $f(\xi)$
satisfies the following differential equation:
\begin{align}
f'' + \frac{1}{2}Ra u_0 f' \xi = 0
\end{align}
where, primes denote differentiation with respect to the similarity variable.
The solution of this equation can be shown to be:
\begin{align}
f(\xi) &= erf(\xi\sqrt{Ra u_0}/2) \label{eqn:Similarity}
\end{align}
giving the dissolution rate at the top boundary as 
\begin{align}
-\frac{\partial C}{\partial y}(y=0) &= \frac{\sqrt{Ra u_0}}{2\sqrt{x}}. 
\end{align}
Integrating this expression from $x=0$ to $L$, we get the total dissolution rate
to be $Nu = \sqrt{Ra u_0 L} = \sqrt{Pe}$, where $Pe = u_0 L^{*}/\phi D$ is the
Peclet number based on the size of the CO$_2$ plume. 

It can be seen that the total dissolution rate under conditions of forced
convection scales with the length of the plume as $\sqrt{L}$. In the free
convection regime however, the total dissolution rate scales linearly with $L$.
We expect the presence of a critical velocity $u_{c}(L)$ such that, for
slow background flows characterized by $u_{0} < u_{c}(L)$, the system will be in
a state of free (mixed) convection, whereas for $u_{0} > u_{c}(L)$, we will
transition to the forced convection regime. 

To determine the critical background flow velocity $u_{c}(L)$, we need an
expression for the mean dissolution rate as a function of Rayleigh number under
free convective conditions. Recent studies have looked at the time-averaged
Nusselt number under free convective conditions and proposed scalings of the
form $Nu \sim Ra^{\alpha}$. \cite{NeufeldDissolution} have used scaling
arguments and experiments to suggest $\alpha = 0.8$, while \cite{Backhaus} have
used experiments to obtain a best-fit exponent of $\alpha = 0.76$. However, the
experiments by both authors were performed under conditions of very high
Rayleigh numbers, of the order of $10^4$ or higher. In this study, we will use a
simpler model given by
\begin{align}
Nu = \frac{Ra L}{40}
\end{align}
which is the well-known relation between Nusselt number and Rayleigh numbers for
steady convection when $Ra$ is close to the critical Rayleigh number
(\cite{Elder1967}). This relation is known to overpredict the Nusselt number for
large $Ra$, but is in good agreement with experimental data for $Ra < 250$.
Since the focus of this manuscript is on aquifers with lower permeabilities, we use this relation as a useful
approximation whose accuracy will be examined \emph{a posteriori}. Requiring the
free and forced convective fluxes to agree at $u_0 = u_c(L)$, we get
\begin{align}
u_c(L) &= \frac{Ra L}{1600}.
\end{align}
We now look at some values for the parameters characteristic of real formations.
We use data from \cite{BachuData}, where a vast amount of information about the
potential sites in western Canada is presented. Using the data from
\cite{PruessSpycher}, we estimate the density increase $\Delta\rho^*$ to be $5
kg/m^3$ (corresponding to a salt mass fraction of 0.1). Assuming $k^* = 10 mD, g
= 10m/s^2, H^* = 50m, \mu^* = 0.5mPa.s, \phi=0.2$ and  $D^* = 1\times 10^{-9}$,
we get the Rayleigh number to be $Ra = 250$. The convective velocity scale is $U_{r} =
k^*\Delta\rho^* g/\mu^* = 3 cm/year.$ Assuming a CO$_2$ plume size of $2km$, we
can compute the critical background flow velocity needed to transition to a forced
convective regime as $u_c^{*} = 20cm/year$ (where the asterisk denotes
dimensional quantity). 
\subsection{Computational Method}
To investigate the transition from free to forced convective flow, we developed
a highly efficient solver for the governing equations. The numerical solution of
forced convection problems becomes computationally very expensive due to the
large linear systems which need to be solved for the velocity field. We have
overcome this difficulty by using an eigenfunction expansion in the vertical
direction using a Fourier sine series.  

The equations to be solved are Eqns.(\ref{eqn:Darcy}-\ref{eqn:ade}) subject to
the initial and boundary conditions mentioned above. The velocity field is
solved obtained using a streamfunction formulation, where the streamfunction
$\psi$ is defined using the relation $\mathbf{u} = \nabla\times\psi
\mathbf{e}_{z}$. Applying the curl operator to Eqn. (\ref{eqn:Darcy}), we get
the following Poisson equation for the velocity field:
\begin{align}
\nabla^{2} \psi & = -\frac{\partial C}{\partial x} \label{eqn:psiPoisson}
\end{align}
The boundary conditions are given by $\psi = 0$ at $y=0$, $\psi = u_0$ at
$y=1$, $\psi = u_0 y$ along the left boundary and $\partial \psi/\partial x = 0$
along the outflow boundary. We now expand the streamfunction in a Fourier
sine-expansion in the vertical direction as:
\begin{align}
\psi(x,y) &= u_{0}y + \sum_{n=1}^{N} \widehat{\psi}_{n}(x) sin(n\pi y)
\end{align}
Substituting this expansion in Eqn. (\ref{eqn:psiPoisson}) followed  by a
multiplication with $sin(m\pi y)$ and integration in the vertical direction, we
get:
\begin{align}
\frac{d^{2}\widehat{\psi}_{m}}{d x^2} - n^2\pi^2\widehat{\psi}_{m} &=
-2\int_{0}^{1}\frac{\partial C}{\partial x}sin(m\pi y) dy \label{eqn:psiODEs}
\end{align}
This solution method depends crucially on a theorem about Fourier transforms
which guarantees convergence of the term-by-term integral of a Fourier
series, even if the original series is divergent (see Theorem 9.8.1 from
\cite{Prosperetti}). This issue is important to consider since the integrand
$\frac{\partial C}{\partial x}$ may not necessarily vanish at the boundaries. 

Eqn. (\ref{eqn:psiODEs}) is now a set of $N_y$ ordinary differential equations
in the horizontal direction, where $N_y$ is the number of vertical grid-points.
These equations can be discretised using the standard second-order finite
difference method, resulting in a set of tridiagonal equations that can be
solved efficiently. Once the velocity field is obtained from the streamfunction,
the concentration field is updated using a second order explicit Adams-Bashforth
method. During the very first time step, we use a second order Runge-Kutte
method for advancing the concentration solution. 

The simulations are performed using 128 uniformly spaced nodes in the vertical
direction and $L\times 128$ uniformly spaced nodes in the horizontal direction.
The variables are located on the grid in a staggered fashion, with
concentrations evaluated at the center of the cells and velocities evaluated on
the faces. The streamfunction is defined at the top-right corner of each cell.
We use constant intervals for timestepping, with a nondimensional timestep of
$\Delta t = 10^{-4}$, chosen to satisfy both the diffusive stability condition
$\frac{1}{Ra}\frac{\Delta t}{\Delta x^2} < \frac{1}{2}$ and the
Courant-Friedrichs-Levy condition $\frac{u_0 \Delta t}{\Delta x} < 1$
(\cite{Ferziger}).

\section{Results}
\begin{figure}
\begin{center}
\includegraphics[width = 0.8\textwidth]{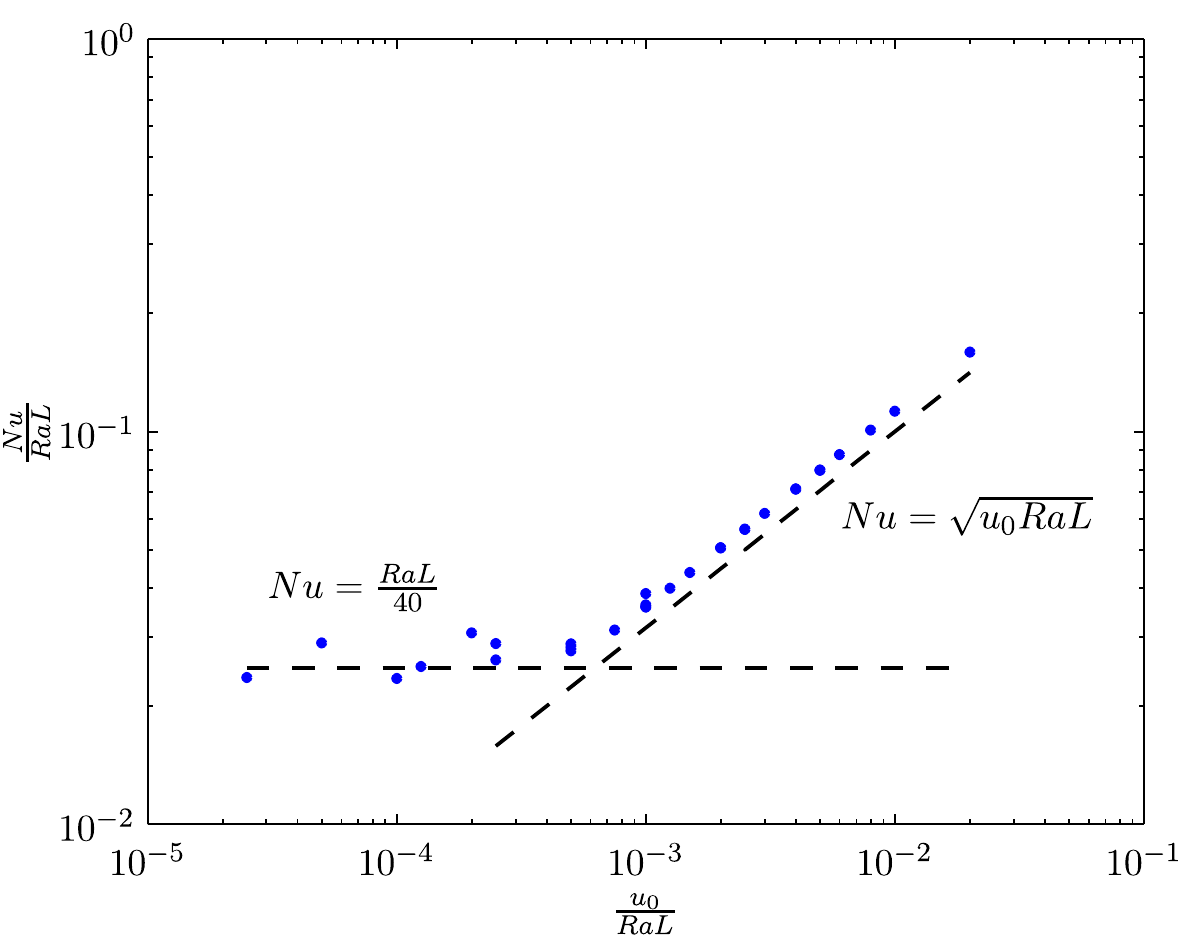}
\vspace{-0.5cm}
\end{center}
\caption{Comparison of scaled Nusselt number, $\frac{Nu}{Ra L}$ computed from
simulations (blue circles) and the theoretical predictions for free and forced convective regimes (dashed
lines) for different background flow rates, scaled as $\frac{u_0}{Ra L}$}
 \label{fig:Nusselt}
\end{figure}

We have used the computational technique described above to investigate the
transition from free to forced convection for $Ra = 50, 100, 200$ and $400$, and
for $u_0$ varying between $0.1$ and $10.0$. The length of the CO$_2$ source was
varied between $L=10$ to $L=40$. In Fig. (\ref{fig:Nusselt}), we plot the
Nusselt numbers computed for the simulations scaled with $RaL$, where the
Nusselt number is defined as:
\begin{align}
Nu &= \int_{x=0}^{L}-\frac{\partial C}{\partial y}(y = 0)dx
\end{align}
\begin{figure}[t]
\begin{center}
\includegraphics[width = 3.3in]{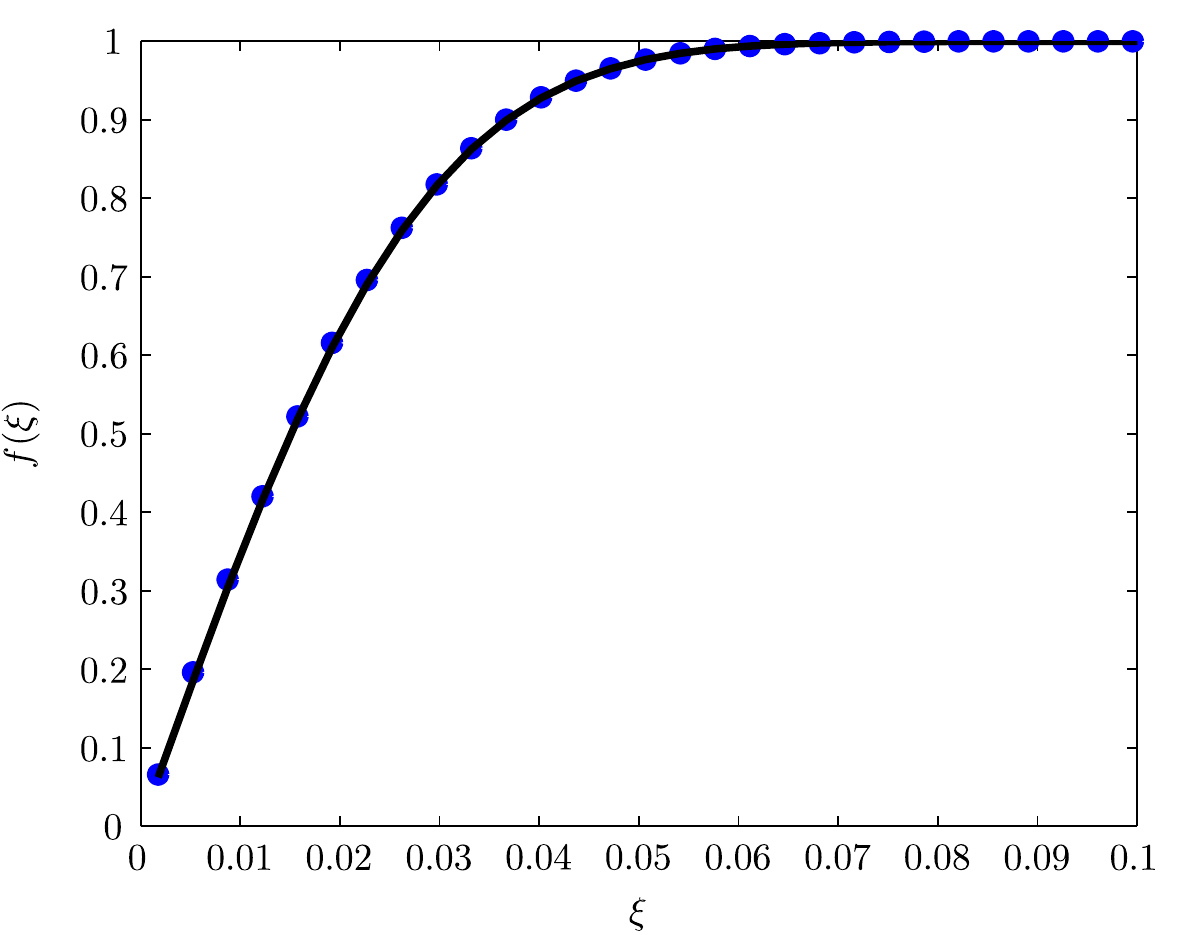}
\vspace{-0.5cm}
\end{center}
\caption{Comparison between the theoretically predicted concentration profile
$C(x,y) = 1 - erf(y\sqrt{Ra u_0}/2\sqrt{x})$ (black line) and that computed from
a simulation (blue symbols)) with $Ra = 400, u_0 = 10$ along a line at the
center of the CO$_2$ source.}
 \label{fig:profile}
\end{figure}

We first verify that the grid spacing used in the simulations is fine enough to
resolve the boundary layer by performing a simulation with $Ra = 400$ and $u_0
= 10$ ( the highest values for both parameters considered in this manuscript).
We compare the profile of dissolved carbon dioxide in the vertical direction at
the center of the plate with the similarity solution given by Eqn.
(\ref{eqn:Similarity}). In Fig. (\ref{fig:profile}), we plot the computed
boundary layer profile using blue symbols and the analytical solution using the
black line. The simulations are seen to match the theoretical solution well,
confirming that the resolution is fine enough to resolve the boundary phenomena.

\begin{figure}
\begin{center}
\includegraphics[width = 0.6\textwidth, angle=-90]{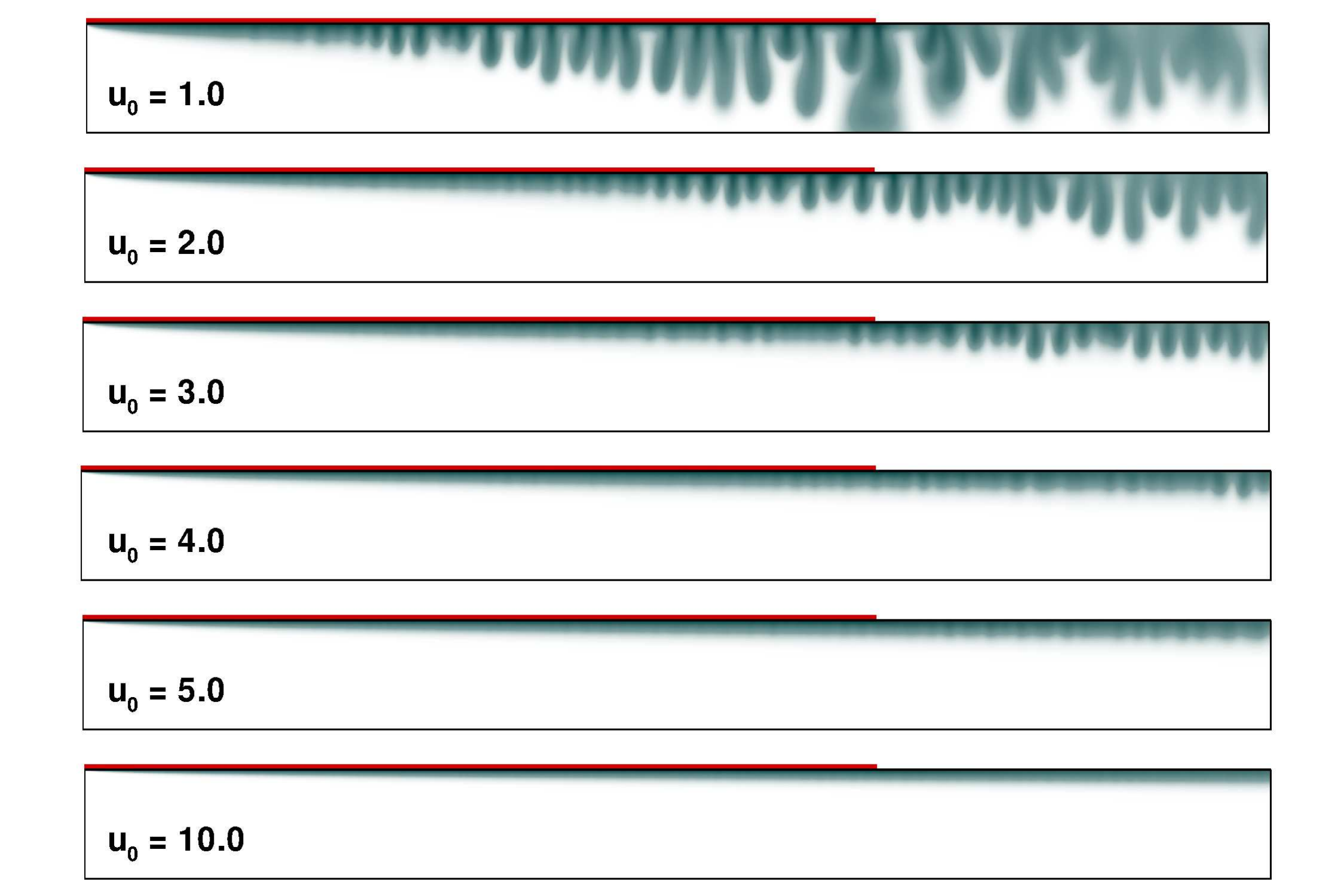}
\vspace{-0.5cm}
\end{center}
\caption{Contours of concentration of dissolved CO$_2$ for simulations with
$Ra = 400$ and $L = 10$. The predicted critical velocity is $u_{c} = 2.5$,
beyond which the dynamics near the CO$_2$ source are predicted to resemble
forced convection. The location of the CO$_2$ source is marked in red.}
 \label{fig:contours}
\end{figure}

\begin{figure}
\begin{center}
\includegraphics[width = 0.6\textwidth]{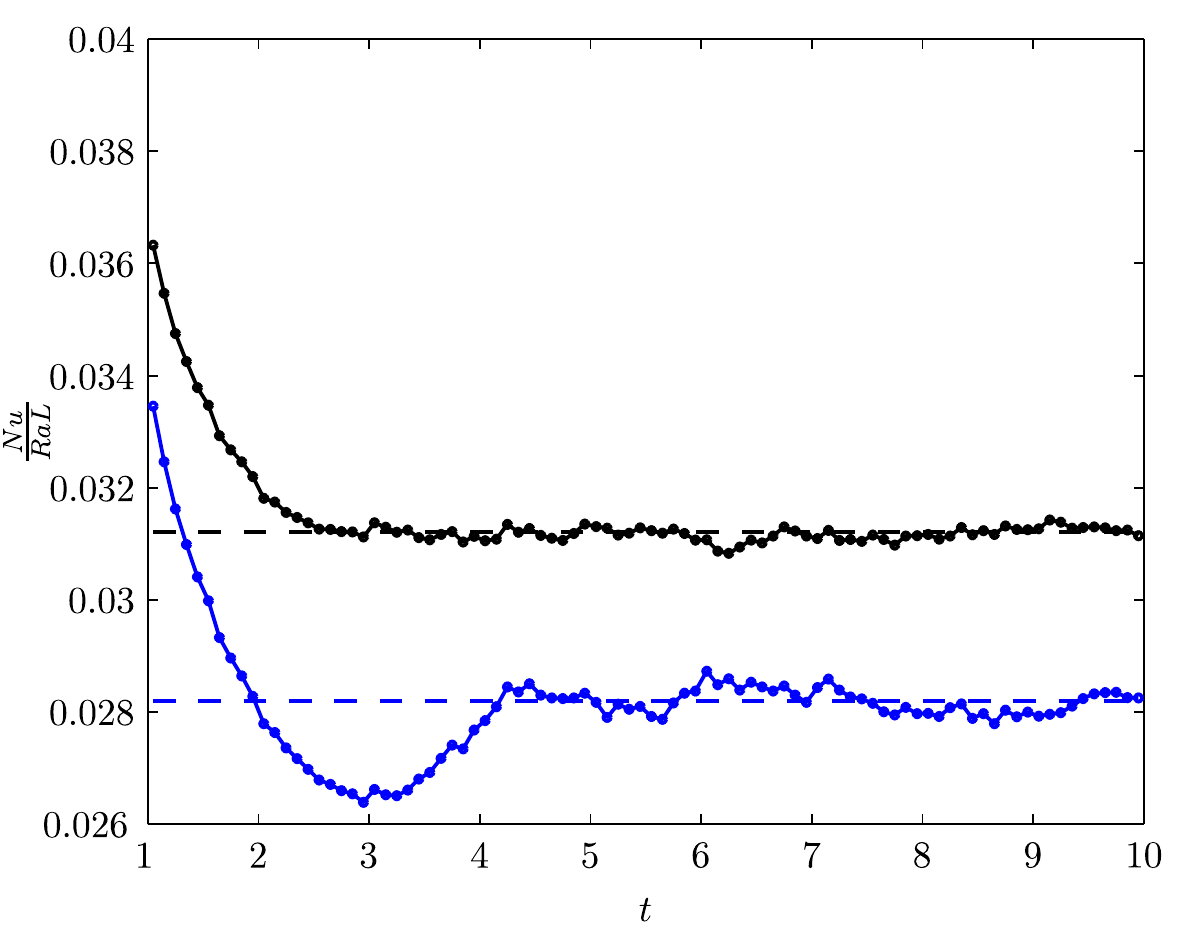}
\vspace{-0.5cm}
\end{center}
\caption{Nusselt number at different times for two simulations with $Ra = 400$
and $u_0 = 2.0$ (blue line) and $u_0 = 3.0$ (black line). The critical velocity
from theory is $u_0 = 2.5$.}
 \label{fig:Nu_vs_t}
\end{figure}

From the scaling arguments presented in the
previous section, $Nu/RaL = 1/40$ in the free convection regime for low Rayleigh numbers and $Nu/RaL =
\sqrt{u_0/RaL}$ in the forced convective regime. These scalings are plotted
using dashed lines in Fig. (\ref{fig:Nusselt}). It must be noted that the
parameter combination $RaL$ that occurs in the analysis is a Rayleigh-number
that is based on choosing the size of the CO$_2$ source as a length scale
instead of the depth of the medium. The simulations are performed upto time
$T=10$ and the Nusselt number is computed by averaging the instantaneous values from $t=9.5$ to $t=10$, 
and these average values are plotted using blue circles. It can be seen that
both the computed Nusselt numbers and the critical velocity agree very well with
those predicted using theoretical arguments.
For concreteness, we present contours of the dissolved concentration of CO$_2$
for simulations fixed with $Ra = 400$ and $L=10$ in Fig. (\ref{fig:contours})
for background flow velocities of $u_0 = 1,2,3,4,5$ and $10$. The predicted
critical velocity for transition to forced convection is $u_c = 2.5$. The
contours show instantaneoous snapshots of the system at $t=10$. It is seen that
for velocities $u_0 > u_c$, the flow resembles forced convective behavior
closely. Also presented in Fig. (\ref{fig:Nu_vs_t}) are the computed $Nu/RaL$ at
different times for this problem with $u_0 = 2$ (blue line) and $u_0 = 3$
(black line). The simulation with $u_0 = 2$ is slightly below the critical
velocity and demonstrates free-convective behavior which can be identified using
the distinct minima in the dissolution rate followed by an enhancement due to
the effect of fingering. The case with $u_0 = 3$ on the other hand achieves a
steady state dissolution rate and does not display the minima. The dashed lines
represent average Nusselt numbers that were used in Fig. (\ref{fig:contours}).
The fluctuations in the Nusselt number are due to the fact that the simulations
are continuously forced (once every 100$\Delta t$), to produce fingering
phenomena when that is the preferred state of the system. In the absence of
continous forcing, the simulations can reach an artificial steady state even
when fingering is the preferred state after the initial perturbations have been
\emph{flushed} out of the domain by the background flow. 

\section{Discussion}
The results presented in this manuscript describe how the presence of a
background flow alters the dynamics of CO$_2$ dissolution in saline aquifers.
The main implications of these results are:
\begin{enumerate}
  \item An accurate understanding of the magnitude and direction of background
  flows is needed to locate monitoring wells. In the presence of a reasonably
  strong background flow, monitoring wells located right under the CO$_2$ plume
  to investigate dissolved carbon content might never record fluid with
  meaningful concentrations of carbon dioxide. In the presence of experimental
  data on existing background flow conditions, these simulations can be used to
  suggest suitable locations to place monitoring wells. 
  \item For low Rayleigh number systems, forced convection can result in far
  higher dissolution rates than free convection. Further, the magnitude of the
  rates of dissolution can be estimated quite accurately using the well-known
  theoretical formulae. Another benefit of the constant dissolution rate due
  to forced convection has to do with upscaling into larger reservoir simulators
  where the boundary layer cannot be resolved. 
  \item In this work, we have not explicitly included the role of either
  hydrodynamic dispersion, or anisotropy of the permeability tensor. In the
  context of the simple theoretical arguments presented in this paper,
  hydrodynamic dispersion essentially serves to define an effective
  (higher) diffusion coefficient which will push the critical velocity lower and
  the Nusselt number under conditions of forced convection higher. Similarly,
  the presence of a permeability anisotropy with higher horizontal permeability
  will tend to make forced convection even more efficient than suggested in this
  manuscript. 
\end{enumerate}
\section{Acknowledgements}
This work was supported by the US Department of Energy through the Zero
Emissions Research and Technology (ZERT) Project.

\bibliographystyle{abbrvnat}
\bibliography{saikiran}
\end{document}